\documentclass[conference]{IEEEtran}
\IEEEoverridecommandlockouts
\usepackage{cite}
\usepackage{amsmath,amssymb,amsfonts}
\usepackage{graphicx}
\usepackage{textcomp}
\usepackage{xcolor}
\usepackage{dblfloatfix}
\usepackage{tabu}                      
\usepackage{booktabs}  
\usepackage{amsmath, xparse}
\usepackage{lipsum}                    
\usepackage{mwe}                       
\usepackage{mathptmx} 
\newcommand{\mathrmbf}[1]{\mathrm{\mathbf{#1}}}
\usepackage{bm}
\usepackage{amsfonts}
\usepackage{amssymb}
\usepackage{stackengine}

\usepackage{amsmath}
\usepackage{paralist}
\usepackage[font={small,sf},caption=false,captionskip=1pt]{subfig}
\usepackage{color, soul}
\usepackage[font={scriptsize,sf},tableposition=top]{caption}
\usepackage{float}
\floatstyle{plaintop}
\restylefloat{table}
\usepackage{siunitx}
\usepackage{colortbl}
\usepackage{sparklines}
\usepackage{icomma}

\usepackage{algorithm} 
\usepackage{algpseudocode} 
\usepackage{gensymb}

\usepackage{hyperref}

\def\BibTeX{{\rm B\kern-.05em{\sc i\kern-.025em b}\kern-.08em
    T\kern-.1667em\lower.7ex\hbox{E}\kern-.125emX}}

\begin{document}

\title{An Incremental Multi-Level, Multi-Scale Approach to Assessment of
Multifidelity HPC Systems\\
}

\author{\IEEEauthorblockN{Shilpika}
\IEEEauthorblockA{
Argonne Leadership Computing Facility\\
Argonne National Laboratory\\
Email: shilpika@anl.gov}
\and
\IEEEauthorblockN{Bethany Lusch}
\IEEEauthorblockA{Argonne Leadership Computing Facility\\
Argonne National Laboratory\\
Email: blusch@anl.gov}
\and 
\IEEEauthorblockN{Venkatram Vishwanath}
\IEEEauthorblockA{Argonne Leadership Computing Facility\\
Argonne National Laboratory\\
Email: venkat@anl.gov}
\and 
\IEEEauthorblockN{Michael E. Papka}
\IEEEauthorblockA{Argonne Leadership Computing Facility\\
Argonne National Laboratory \& \\
Department of Computer Science \\
University of Illinois Chicago\\
Email: papka@anl.gov}
}

\maketitle

\begin{abstract}
With the growing complexity in architecture and the size of large-scale computing systems, monitoring and analyzing system behavior and events has become daunting. Monitoring data amounting to terabytes per day are collected by sensors housed in these massive systems at multiple fidelity levels and varying temporal resolutions. In this work, we develop an incremental version of multiresolution dynamic mode decomposition (mrDMD), which converts high-dimensional data to spatial-temporal patterns at varied frequency ranges. Our incremental implementation of the mrDMD algorithm (I-mrDMD) promptly reveals valuable information in the massive environment log dataset, which is then visually aligned with the processed hardware and job log datasets through our generalizable rack visualization using D3 visualization integrated into the Jupyter Notebook interface. We demonstrate the efficacy of our approach with two use scenarios on a real-world dataset from a Cray XC40 supercomputer, Theta.
\end{abstract}


\section{Introduction}
\label{sec:introduction}
High-performance computing (HPC) systems are designed to address some of the most intricate computational challenges in research, solving complex problems, and driving discoveries and innovations in real-time. These systems operate at speeds exceeding the fastest commodity desktops or servers by quadrillions of calculations per second, emphasizing the urgency and importance of their real-time operations. 
The supervision and monitoring of supercomputer systems are critical for ensuring their robustness and trustworthiness. As these systems evolve in complexity with changing execution environments, system architectures, increasing computation capabilities, and frequent software/hardware updates and upgrades, the monitoring data collected also increases. With the advent of newer and more complex artificial intelligence models, applications and jobs that utilize these systems may require months to execute, making any system errors and failures a potential threat to the integrity of the results, leading to inefficient use of resources and adding significant overhead to the research and development process. 

In order to maintain robust and efficient HPC systems, we analyze the log data, which can be used for various tasks, including debugging, anomaly detection, and performance estimation. Consequently, we handle multifidelity, large-scale data from various components housed within the HPC system. For instance, the hardware error log includes information from diverse and interconnected control systems and subsystems, with data sizes reaching tens of gigabytes (GB). The environmental log data, collected at intervals of 10-30 seconds, accumulate in the gigabyte (GB) to terabyte (TB) range every few weeks. It primarily constitutes data from system sensors at the rack, blade, and node level, e.g., power consumption, air and water temperatures, and voltages at various sections in the system. Additionally, the job log data detailing the applications utilizing the systems and their attributes (e.g., nodes used, start and end times) accumulates to hundreds of megabytes (MB) annually. Analyzing these extensive datasets, averaging approximately terabytes in a few weeks or even days, presents a significant challenge. Therefore, there is a dire need to process these logs in a timely and prompt manner to enable fast research through more efficient systems.

We built an online analytical system that incrementally processes this large log data, specifically processing the (i) environment logs (time-series data). This environment dataset is massive, with a sampling rate typically ranging in seconds, and holds information regarding the current state of the system components. We then visually align the extracted patterns with (ii) job logs (numerical and text data) and (iii) hardware logs (numerical and text data) logs. Therefore, we deal with multifidelity data from various origins within the HPC system. 

The work utilizes multiresolution Dynamic Mode Decomposition (mrDMD) to concisely represent the more extensive environment log data. The results are then compared with additional preprocessed log data, such as hardware and job log data. Dynamic Mode Decomposition (DMD) decomposes the data into spatial modes and corresponding time dynamics for each mode. Unlike its predecessor, Dynamic Mode Decomposition (DMD), mrDMD recursively subtracts low-frequency, or slower varying, dynamics from the data at each selected timescale, making it ideal for separating different timescales for analysis. mrDMD also eliminates high-frequency noise at each timescale and can reduce the data size from terabytes to megabytes. Using these extracted modes, we can reconstruct the original data without the high-frequency noise.

Previous works have employed the DMD algorithm to isolate and extract distinct sleep spindle networks from sub-dural electrode array recordings of human subjects using DMD spectrum analysis with baseline analysis~\cite{BRUNTON20161}. 
 The mrDMD extracts the spatial and temporal features over multiple timescales by integrating space and time. In our previous work, a version of mrDMD was applied to HPC log datasets, using the mrDMD spectrum and baselines representing normal behavior \cite{shilpika2023multilevelmultiscalevisualanalytics, shilpika2023visual}. However, we found that although the mrDMD analysis with the spectral decomposition helps extract patterns of interest at multiple timescales and frequency ranges, it needed to be more scalable given the massive size of the environment log dataset. Therefore, in this work, we present an online version of the mrDMD with spectral analysis to process the environment logs in a timely and prompt manner for streaming data. We extract the high-power mrDMD modes at various frequency ranges incrementally. We then choose baselines representing expected system behavior to extract patterns that indicate changes between the current and baseline states.
This paper answers the following questions to enhance resilience in HPC systems:
\begin{itemize}
\item[Q1]  Are the extracted mrDMD modes reliable enough to represent the underlying system dynamics?
\item[Q2] What is the difference in  accuracy between online and regular mrDMD?
\item[Q3] Does the system behavior extracted from the environment logs correlate with faults seen in hardware and job failures? If so, what patterns can be extracted, and what is the computational cost of the analysis?
\end{itemize}

We make the following contributions:
\begin{itemize}
\item We develop a new incremental algorithm to generate the mrDMD modes (I-mrDMD), which enables fast analyses of time series data collected from online data streams.
\item We develop a general visual analytics solution to represent multiple supercomputer racks based on a set of parameters detailing the supercomputer layout. We achieve this using D3~\cite{D3} integrated into Jupyter Notebook~\cite{jupyter}.
\item We apply our solution to two case studies using real-world supercomputer log data to drill down to the supercomputer usage and visually align the behaviors across environment, job, and hardware logs.
\end{itemize}
\section{Related Work}
\label{sec:relatedwork}

\subsection{Log Analysis in Large Scale Computing Systems}
Significant efforts have been devoted to debugging and forecasting failures in large-scale HPC systems, enabling proactive measures for failure detection. Previous survey papers~\cite{ELMASRI2020106276, surveyreliability, 7412635, b32,b33} on log diagnosis, forecasting, abstraction, and failure classification provide a comprehensive overview of current methodologies, highlighting their significance and limitations in log data analysis. 

Several machine learning, deep learning, and transformer-based approaches have recently gained popularity, such as a generative adversarial network based in long short-term memory (LSTM) for anomaly detection using permutation event modeling~\cite{LOGGAN}, a Convolutional Neural Networks-based model to learn event relationships and automatically detect anomalies~\cite{8511880}, and a self-supervised framework using Bidirectional Encoder Representations from Transformers (LogBERT)~\cite{guo2021logbertloganomalydetection}. Earlier log analysis solutions include coordinated views for network topology aimed at optimizing network communication and system behavior in supercomputers~\cite{ FUJIWARA201898, 8048931, shilpika2019mela, 9825952,shilpika2023multilevelmultiscalevisualanalytics, 9825952, 8585646, 8585646}, anomaly detection frameworks with micro-services architectures using execution logs and query traces~\cite{8957683},
and functional data analysis (FDA) to incrementally and progressively compute the online data streams of environment logs~\cite{9751445}.  
Some proposed visual analytic approaches to analyze millions of inconsistent text records and attribute structure extraction use information theory~\cite{255601, 10.1145/3411764.3445396}. Others employ signal processing, pattern mining and recognition, and text and event-based correlations~\cite{b2,b13,b14,b15,b16}. Online analysis of log data is crucial to promptly and succinctly represent numerical and textual data, capturing the evolving behavior of the system in a timely and efficient manner~\cite{9139847, 8813251, 9751445, b31, 6903589, 9209707}. Although our analysis mainly focuses on environment logs, our visualization can display filtered and curated results from multiple logs (hardware log, job log, and environment log data) to give a succinct representation of the supercomputing system, and it does so promptly in an online fashion. 

\subsection{Dynamic Mode Decomposition and its Multiresolution Counterpart}

As mentioned above, mrDMD is a counterpart of the DMD algorithm for modeling multiscale spatiotemporal systems. The mrDMD algorithm was initially applied to foreground and background subtraction in video feeds~\cite{mrDMD}. The DMD algorithm approximates the modes of the Koopman operator from dynamical systems theory and decomposes spatiotemporal data into dynamic modes derived from temporal measurements~\cite{mrDMD,fd_mrdmd,schmid_2010}. Building on DMD, the mrDMD method recursively processes timescales from the nonlinear dynamical system, selecting dynamic modes to remove at each scale. This approach was adapted from multiresolution analyses such as wavelet methods and windowed Fourier transforms~\cite{mrDMD,shilpika2023multilevelmultiscalevisualanalytics}. 

In recent years, both DMD and mrDMD have served as effective tools for investigating the nonlinear dynamics of systems such as complex flows modeling~\cite{schmid_2010, mrDMD, tut}, wind turbines damage detection~\cite{windTurbine}, airplane wing surface pressure~\cite{GONZALES2022107718} infectious disease monitoring~\cite{cntrl} sleep spindle clustering~\cite{BRUNTON20161}, streaming analysis~\cite{GONZALES2022107718, onlineDMD, Hemati2014DynamicMD, Pendergrass2016StreamingGS}, supercomputer log analysis~\cite{shilpika2023multilevelmultiscalevisualanalytics}, and foreground and background separation in video analysis~\cite{mrDMD}. 
Some works extend the mrDMD method to apply to a wider variety of multiscale systems and reconstruct the input using the extracted  components~\cite{PhysRevE.99.063311}. In our work, we adopt the methodology previously applied to supercomputer environment logs~\cite{shilpika2023multilevelmultiscalevisualanalytics}. However, this method lacks scalability when applied to large volumes of data. To counter this problem, we developed an incremental streaming version of the pipeline proposed previously~\cite{shilpika2023multilevelmultiscalevisualanalytics}.

Several methods have been proposed for streaming data analysis using DMD~\cite{onlineDMD, Hemati2014DynamicMD, Pendergrass2016StreamingGS} and mrDMD~\cite{GONZALES2022107718}. As our work builds upon previously developed mrDMD approaches~\cite{shilpika2023multilevelmultiscalevisualanalytics, mrDMD, 10.1063/1.4863670, humaticlabs}, other DMD streaming methods are beyond the scope of this paper. Joseph et al. proposed a windowed approach for mrDMD~\cite{GONZALES2022107718} to update mrDMD results rather than recompute from scratch. The windowed approach converts data into smaller overlapping time series, resulting in data reconstructions that are staggered in nature. The work combines non-overlapping regions and trusting the data that best matches the overlapping regions. In contrast, our work employs the more recently developed incremental SVD~\cite{KUHL2024109022} to incrementally update the DMD mode decomposition at each scale of the multiresolution step, eliminating overlaps. 
\section{Methodology}
\label{sec:methodology}
In this section, we describe the set of adaptations of the mrDMD method that promptly generate the spatiotemporal mrDMD modes. In section \ref{sec:methodology_mrdmd}, we summarize the mrDMD algorithm. We then detail our improvements necessary to apply the mrDMD to the massive streaming environment logs. Following this, we discuss our generic visualization for computer rack display and interactions. The code is available at \url{https://github.com/sshilpika/incremental-mrdmd}.

\begin{figure}[h]
	\centering
    \includegraphics[width=\linewidth]{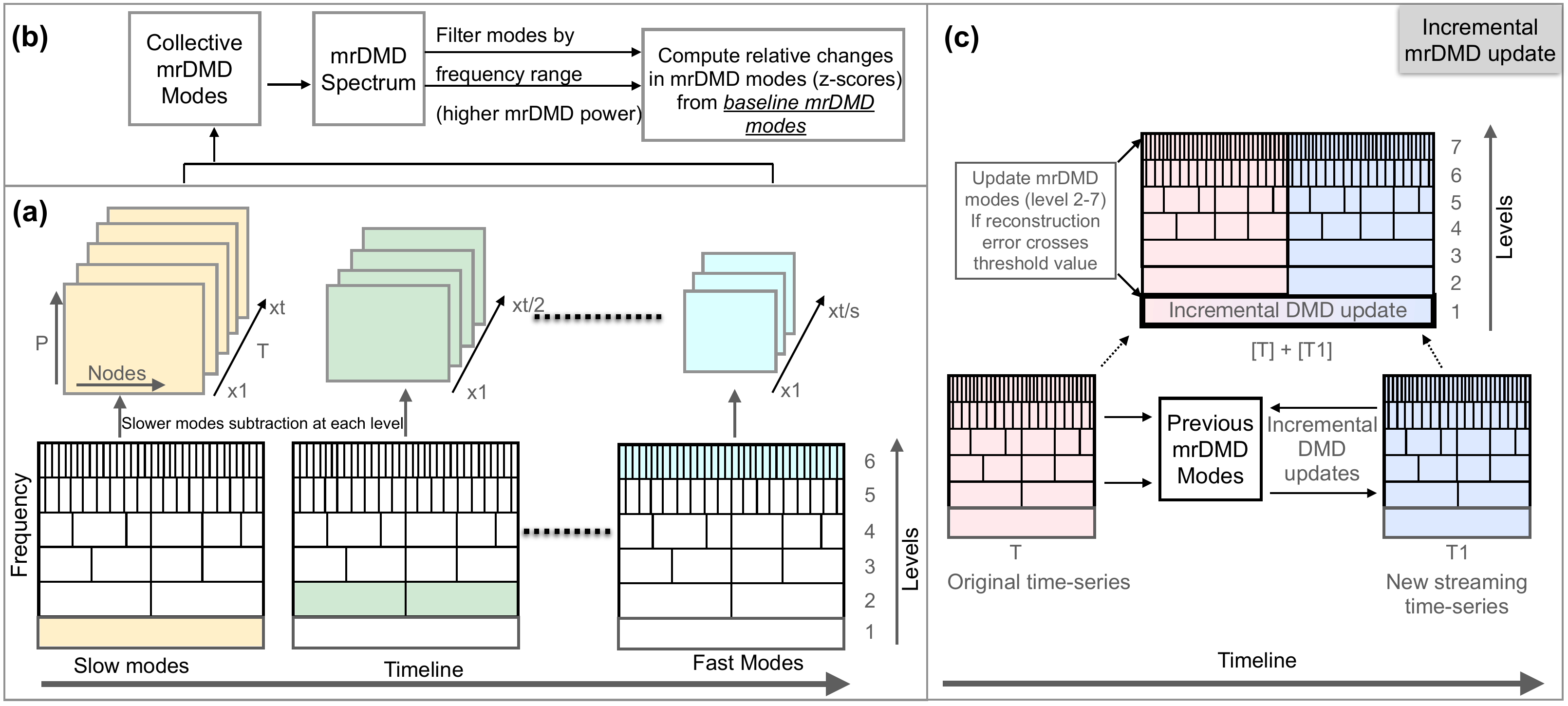}
    \caption{Figure shows (a) Multiresolution Dynamic Mode Decomposition (mrDMD), (b) frequency Isolation of spatiotemporal modes using
mrDMD spectrum, and (c) Incremental Multiresolution Dynamic Mode Decomposition (I-mrDMD)}
	\label{sys_arch}
\end{figure}

\subsection{Multiresolution Dynamic Mode Decomposition}
\label{sec:methodology_mrdmd}

Consider environment log measurements (sensor readings) taken from $P$ sensors at $T$ snapshots in time sampled at every $\Delta t$. We construct two largely overlapping raw data matrices, $\mathrmbf{X}$ and $\mathrmbf{Y}$, such that the elements of $\mathrmbf{Y}$ are shifted by $\Delta t$ from those in $X$, as shown below. \\

$
    \mathrmbf X = 
        \begin{bmatrix}
            | & | & | & & | \\ 
            \mathrmbf{x}_{1} & \mathrmbf{x}_{2} & \mathrmbf{x}_{3} & ... & \mathrmbf{x}_{T-1} \\
            | & | & | & & |
            
        \end{bmatrix}
$ \\
\newline

$
    \mathrmbf Y = 
        \begin{bmatrix}
            | & | & | & & | \\ 
            \mathrmbf{x}_{2} & \mathrmbf{x}_{3} & \mathrmbf{x}_{4} & ... & \mathrmbf{x}_{T} \\
            | & | & | & & | 
            
        \end{bmatrix}
$ \\

The DMD algorithm indirectly solves for the unknown linear operator, also known as the best-fit matrix $\mathrmbf{A}$, such that

\begin{equation}
\label{eq:yax}
    \mathrmbf{Y} = \mathrmbf{AX} 
    \implies \mathrmbf{A} = \mathrmbf{YX^\dagger}   
\end{equation}
where $\dagger$ is the Moore–Penrose pseudoinverse. Dynamic mode decomposition (DMD) is the simply the eigendecomposition of this best-fit matrix $\mathrmbf{A}$. Here $\mathrmbf{A}$ is chosen to minimize $ \Vert x_{t+1}- Ax_{t} \Vert_2$ over the  $t = 1,2,3,...,T-1 $ snapshots. The approach used in this work to compute this approximation of $\mathrmbf{A}$ is to employ the singular-value decomposition (SVD) of the data matrix $\mathrmbf{X}$.
Given that the values of P can be exceedingly large, particularly when dealing with supercomputer sensors, the eigendecomposition of $\mathrmbf{A}$ becomes computationally expensive. Positing the existence of an inherent low-dimensional spatial structure within X, the following algorithm enables the calculation of DMD modes and eigenvalues without requiring direct computation of the matrix $\mathrmbf{A}$~\cite{Brunton_Kutz_2019}:\\
1) First, evaluate the SVD~\cite{numlin} of the data matrix $\mathrmbf{X}$:
    \begin{equation}
    \label{eq:svd}
    \mathrmbf{X} = \mathrmbf{U}\boldsymbol{\Sigma} \mathrmbf{V^{'}}
    \end{equation}
where $'$ is the conjugate transpose, $\mathrmbf{U}\in \mathbb{C}^{P\times r}$ , $\boldsymbol{\Sigma} \in \mathbb{C}^{r\times r}$ and $\mathrmbf{V}\in \mathbb{C}^{T-1\times r}$. Here $r$ is the reduced rank of the SVD approximation of $\mathrmbf{X}$ computed using the optimal Singular Value Hard Threshold (SVHT~\cite{opsvht}).\\
2) $\tilde{\mathrmbf{A}}$ is the $r\times r$ projection of the full matrix $\mathrmbf{A}$ onto the low-rank modes of $\mathrmbf{U}$. Substituting (\ref{eq:yax}) and (\ref{eq:svd}),
\begin{equation}
\label{eq:atilde}
\tilde{\mathrmbf{A}} = \mathrmbf{U^{'}AU} = \mathrmbf{U^{'}YV}\boldsymbol{\Sigma}^{-1}
\end{equation}
\\
3) Solve for the eigendecomposition of $\tilde{\mathrmbf{A}}$ :
\begin{equation}
\label{eq:eigdecomp}
\tilde{\mathrmbf{A}}\mathrmbf{W = W}\boldsymbol{\Lambda}
\end{equation}
where the columns of $\mathrmbf{W}$ are the eigenvectors and the corresponding eigenvalues  $\mathrm{\lambda}_i$  are along the diagonal of the diagonal matrix $\boldsymbol{\Lambda}$.\\
4) Finally, reconstruct the eigendecomposition of $\mathrmbf{A}$ from $\mathrmbf{W}$ and $\boldsymbol{\Lambda}$. 
\begin{equation}
\label{eq:eigdecomp1}
\boldsymbol{\Phi} = \mathrmbf{YV}\boldsymbol{\Sigma}^{-1}\mathrmbf{W}
\end{equation}
Each column of $\boldsymbol{\Phi}$, $\mathrm{\phi_i}$, contains a DMD mode or eigenvector of $\mathrmbf{A}$ which corresponds to the $ith$ eigenvalue in $\boldsymbol{\Lambda}$.

Once the DMD modes are computed, the projected future results are obtained as follows~\cite{Brunton_Kutz_2019}:
\begin{equation}
\mathrm{\tilde{y}(t) = \sum_{i=1}^{r} \phi_{i}\hspace{0.1cm}exp(\psi_i t)}\hspace{0.1cm}a_{i(0)} = \boldsymbol{\Phi}diag(exp(\psi t))\boldsymbol{a}
\label{eq:o}
\end{equation}
where we have rewritten $\mathrm{\psi_i = ln(\lambda_i)/\Delta t}$, with data sampled every $\Delta t$, $diag(exp(\psi t))$ is the eigenvalue diagonal matrix $exp(\psi t)$, \textit{r} is the target rank of SVD, $a_{i(0)}$ is the initial amplitude of each mode, and $\boldsymbol{a}$ is a vector of the coefficients $a_i$.

The mrDMD is an improvement to the Dynamic Mode Decomposition (DMD) algorithm~\cite{mrDMD} created to screen lower to higher-frequency dynamics from temporal data recursively, capturing the system's dynamics at varying time scales. At each level (Fig.~\ref{sys_arch}(a)-(yellow, green)) the DMD modes are processed. The first level processes the entire timeline and extracts the slowest DMD modes, i.e., those with frequencies below a specified threshold. We compute this threshold by computing the desired number of oscillations per timestep. For level 2, the timeline of the data (reconstructed by subtracting out the slower modes) is split at specific timesteps. For example, in Fig.~\ref{sys_arch}(a)-(green), the data is split into halves. Each split is simultaneously processed to extract slower modes but at a finer temporal resolution. We then repeat the steps until we reach the termination criterion, for example, the maximum level. This recursive computation prevents DMD modes from dominating any frequency range in the final result. Once the mrDMD modes are computed, the projected future results are obtained as in Eq.~\ref{eq:o} and is given by~\cite{mrDMD, Brunton_Kutz_2019}:

\begin{equation}
\mathrm{x_{mrDMD}(t) = \sum_{i=1}^{T} \phi_i^{(1)} \hspace{0.1cm}exp(\psi_i t)\hspace{0.1cm} a_{i(0)}}
\end{equation}

\begin{equation}
\mathrm{= \sum_{i=1}^{t_1} \phi_i^{(1)} exp(\psi_i t)\hspace{0.1cm} a_{i(0)} + 
\sum_{i=t_1+1}^{T} \phi_i^{(1)}\hspace{0.1cm} exp(\psi_i t)}\hspace{0.1cm}a_{i(0)}, 
\label{eq:mr}
\end{equation}
 \indent \indent \indent \indent \textit{(slower modes)} $\qquad \qquad \qquad $\textit{(faster modes)} \\

where $\phi_i^{(1)}(x)$ represents the mrDMD modes for $T$ time steps. The first term in Eq.~\ref{eq:mr} represents the slower dynamics of the data, and the second term represents the rest of the dynamics to be recursively extracted at each level~\cite{mrDMD, Brunton_Kutz_2019,shilpika2023multilevelmultiscalevisualanalytics}. As we know, the slower dynamics are extracted at the lower levels. We are then able to sample the data at low frequencies. As we progress towards the higher levels, we gradually increase the sampling rate, thus processing data at finer resolutions in time. We follow the choice of sampling rate from the previous work using mrDMD on the supercomputer logs~\cite{shilpika2023multilevelmultiscalevisualanalytics, shilpika2023visual}, which set the sampling rate to four times the Nyquist limit to capture cycles~\cite{nyq}. The sampling rate is customizable.

\subsubsection{Incremental mrDMD (I-mrDMD) updates}
\label{sec:inc-mrdmd}
\begin{algorithm}
	\caption{Incremental mrDMD} 
 \label{algo1}
	\begin{algorithmic}[1]
        \State $level=1$
        \If {I-mrDMD}
        \State compute a Spatially Parallel / Temporal Serial incrementally updated truncated $rank$-$q$ SVD of $\mathrmbf{X}$~\cite{KUHL2024109022}
        \Else
        \State truncated SVD of $\mathrmbf{X}$~\cite{numlin}
        \EndIf
        
        \For {$previous\_nodes=1,2,\ldots$ L}
            \State $node\_level\leftarrow node\_level + 1$
        \EndFor 
        
        \State $dmd\_modes \leftarrow$ extract slower DMD modes
        \State $data\_recon \leftarrow$ reconstruct time-series with slower modes          
        \State $data \leftarrow data  - data\_recon$
        \State save DMD modes for current level
        \State \textit{split} the data along $T$
        \State $level\leftarrow level + 1$
        \State repeat step $2-11$ for each \textit{split}
	\end{algorithmic} 
\end{algorithm}

The primary bottleneck in the analysis pipeline is the computation of mrDMD spatiotemporal modes for large time series data. Although selecting an appropriate sampling rate at each mrDMD step can reduce the compute time, the analysis remains computationally intensive for large datasets like environmental logs. To address this issue, we implemented an incremental update to the Singular Value Decomposition (SVD)~\cite{KUHL2024109022} step at each level, allowing us to update previously generated SVD results with new data. This approach significantly reduces the compute time required to generate new mrDMD modes for incoming data streams.
For instance, after processing $T$ time steps of data, we perform an incremental SVD update at level 1 for $T1$ new incoming time steps (refer Fig.~\ref{sys_arch}(c)). This update is applied to the previously generated SVD results at level 1, resulting in a new level 1 that includes $T+T1$ time steps. Subsequently, one can update the prior mrDMD results from levels 2 through L, which initially covered the $T$ time steps, based on the difference between the newly computed slower modes and the previous slower modes. 
If the Frobenius norm of this difference exceeds a user-defined threshold, users can asynchronously recompute the mrDMD modes for the original $T$ time steps. To address $Q2$, we conducted experiments to compute the difference between the reconstruction and the actual data for I-mrDMD and mrDMD and found that the difference increases only by a sum of $10$-$5000$, depending on the underlying dynamics of the systems and the time step upgrades. This error is still small when processing timelines in weeks or months, but can accumulate for data processed for multiple months to years. So, we leave this step for future work on different datasets. Since this update for levels 2 through L for previous mrDMD results is an embarrassingly parallel problem, it would not add an overhead to the current computation (Fig.~\ref{sys_arch}(c)).
Finally, we increment the levels for the mrDMD results (Fig.~\ref{sys_arch}(c)-purple), ensuring that the new level 1 contains the $T+T1$ time steps, while the new level 2 corresponds to timeline that is split at $T$ (Fig.~\ref{sys_arch}(c), Algo.\ref{algo1}).

\subsubsection{Frequency Isolation of spatiotemporal modes using mrDMD spectrum}

Previous works have computed DMD~\cite{BRUNTON20161} and mrDMD~\cite{shilpika2023multilevelmultiscalevisualanalytics} power spectrum to further isolate modes of significance from the computed DMD or mrDMD modes. These extracted modes are then compared with the modes from a baseline set by the standard deviation of current minus baseline mode magnitude. The z-scores of this change from baseline were computed for each measurement based on the estimate of standard deviation~\cite{BRUNTON20161}.

For mrDMD spectrum, we compute the frequency of oscillation for $\mathrm{\phi_i}$ in cycles per second as~\cite{BRUNTON20161}: 
\begin{equation}
\label{freq_spec}
\mathrm{f_i = \Bigg  | \frac{imag(\psi_i)}{2\pi}\Bigg |}
\end{equation}

and the mrDMD ``power'' as:
\begin{equation}
\mathrm{P_i = \big  | \phi_i\big |_2^2}
\end{equation}

where $\mathrm{\psi_i = ln(\lambda_i)/\Delta t}$; the imaginary component of $\mathrm{\psi_i}$ determines the frequency of oscillations and the real component determines if the corresponding mode dynamics are growing (positive), or decaying (negative). The mrDMD spectrum, is visualized by plotting the “power” of each mode $\mathrm{\phi_i}$ against its frequency of oscillation $\mathrm{f_i}$ (refer Fig.~\ref{sys_arch}(b) and Sec.~\ref{sec:use_scenarios}). 

\subsection{Visualization}
\label{sec:rack_vis}
\begin{figure}[h]
	\centering
    \includegraphics[width=1\linewidth]{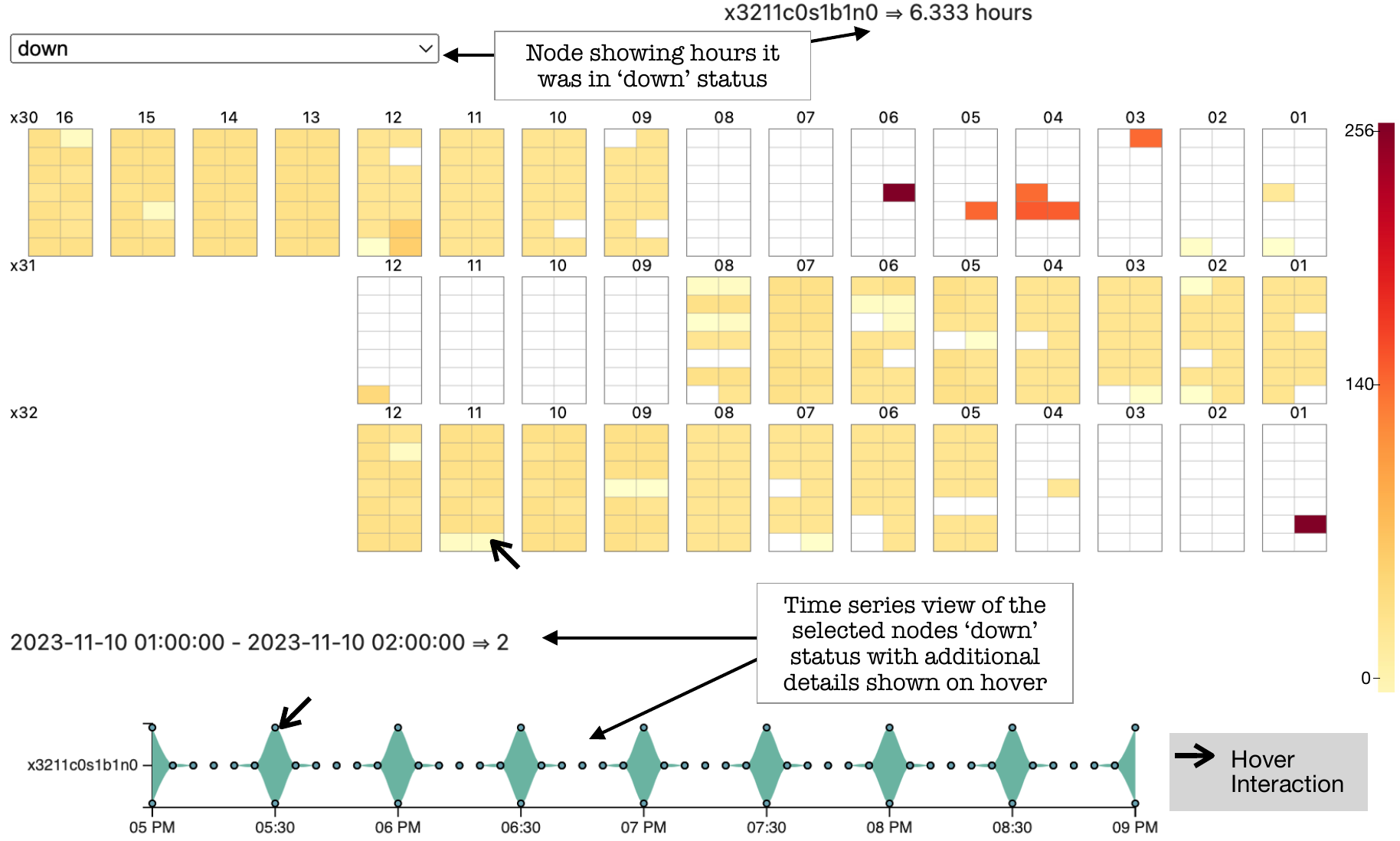}
    \caption{Generalizable rack layout view using D3 Visualization in Jupyter Notebook interface for the supercomputer Polaris~\cite{polaris}}
	\label{viz_first}
\end{figure}

Integrating D3 visualization~\cite{D3} into Jupyter Notebooks~\cite{jupyter} offers several benefits. Firstly, D3's powerful data-driven approach enables the creation of highly interactive and dynamic visualizations, allowing for real-time data exploration and manipulation. This capability is especially valuable in Jupyter Notebooks, where researchers and analysts can seamlessly combine data, code, and visualizations in a single interactive document, enhancing the convenience and efficiency of the process. Additionally, D3's flexibility in handling various data formats and its extensive customization options enable users to create tailored visual representations that can highlight specific insights. This enhances the comprehensibility and presentation of complex data, facilitating a deeper understanding and more effective communication of analytical results.
In this work, we use D3 to generate rack layout views of the supercomputer. The attribute that our visualization uses is a string containing specifications about the layout. 

The format of the string is as follows: ``system\_name rack\_row-alignment rack\_column-alignment Rows[rack-range]:[rack-number-range-per-rack] cabinet\_row-alignment cabinet\_column-alignment Cabinets/Cages:[cabinet-number-range-per-cabinet] slot\_row-alignment slot\_column-alignment Slots:[slot-number-range-per-slot] blade\_row-alignment blade\_column-alignment Blades:[blade-number-range-per-blade] Nodes:[nodes-number-range-per-node]". The row and column alignment takes numbers -1, 1, 2 for right-to-left, left-to-right, and 2 for bottom-to-top (default is top-to-bottom) alignment. For example: ``xc40 1 2 row0-1:0-10 2 c:0-7 1 s:0-7 1 b:0 n:0" is an XC40 system with two rows (0 and 1), eleven racks per row, rows are aligned left-to-right and bottom-to-top, eight cabinets aligned bottom-to-top, eight slots aligned left-to-right one blade aligned left-to-right and one node per blade (refer Fig.~\ref{viz_first},\ref{cs_1_notebook}, and \ref{cs_2_2}). The rack layout view shown in Fig.~\ref{viz_first} has a drop-down selection to isolate different categories of data, such as temperature, voltage, or node statuses. It also has a hover interaction that displays the name of the node. On clicking the node, the time-series data is shown with additional hover interaction to view values within the time-series. In the Fig.~\ref{viz_first}, we display the node down time of the supercomputer for a duration of a few months.

\section{Performance Evaluation}
\label{sec:performance_evaluation}

Here, we evaluate the performance of our algorithms for two example cases. In each case, we simulate a practical streaming analysis context by introducing new time points derived from real-world datasets, ensuring the practicality and relevance of our evaluation.
For the experiments, we used a node of the 560-node HPE Apollo 6500 Gen 10+ supercomputer, Polaris~\cite{polaris}. The node features a 2.8 GHz 7543P processor, 32 AMD Zen 3 (Milan) cores, 512 GB of DDR4 RAM, four NVIDIA A100 GPUs, and two 3.2 TB local SSDs. 
Completion times are averaged over $10$ executions.

\noindent\textbf{Evaluation using supercomputer environment logs.}
Supercomputer logs containing readings of voltages, current, temperatures (water/air/CPU), and fan speeds are collected from various detectors housed in the compute racks (e.g., system boards and coolers). 
Studying these logs to identify failure patterns can aid in making large-scale machines more robust.
There have been several efforts to analyze large-scale computing system logs with visual analytics tools~\cite{guo2018valse, 9751445,shilpika2019mela,tfujiwara2020,cloudet,ensemblelens, shilpika2023multilevelmultiscalevisualanalytics}. 
Our logs (analyzed in \ref{sec:methodology}) collected from a Cray XC40 supercomputer, Theta~\cite{ThetaANL}, contains data from 4,392 nodes housed in 24 racks, with 150 sensor readings per node collected every $15-30$ second intervals.

We evaluate our incremental algorithms on this log data, which encompasses terabytes of information accumulated over a few weeks.
We assume that we already have the temperature readings of size $4,392\!\times\!50,000$ (i.e., $~17$ days).
Then, we add $5,000$ newly arrived time points to this existing data. 
While the update of the I-mrDMD without our incremental update (i.e., recalculation on $55,000$ time points) is finished in $80.580$ seconds, the incremental addition of a time point is completed in $14.728$ seconds.  
Without our incremental approach, updates exceeded the data collection interval, leading to inefficient use of computing resources. Note that we set the \texttt{max\_levels}=$8$ in each case.

This proof-of-concept experiment demonstrates that our improvement is suitable and highly effective for prompt updates of large datasets. 
In practice, the operation experts may select a smaller dataset (e.g., a few hours instead of days), and the ordinary mrDMD may update the result within the current update interval (i.e., every 30 seconds). 
However, most recent supercomputers collect sensor data at $0.03$--$10$Hz~\cite{fugaku,osti_1562918,10.1109/MSPEC.2022.9676353} (i.e., about every 0.1--10 seconds).
Our approach can process large volumes of data to identify anomalies in these large-scale systems promptly.

\vspace{5pt}
\noindent\textbf{Evaluation with GPU metrics data.}
The I-mrDMD is a data-driven approach that enables us to monitor time series collected from various hardware systems. 
We analyze data collected from a 560-node HPE Apollo 6500 Gen 10+ supercomputer, Polaris~\cite{polaris}, as another use scenario. Each node features a 2.8 GHz AMD EPYC Milan 7543P 32-core CPU, 512 GB of DDR4 RAM, and four NVIDIA A100 GPUs interconnected via NVLink. The nodes are also equipped with a pair of Slingshot 11 network adapters and a pair of local 1.6 TB SSDs in RAID 0 configuration for user data storage. Each compute node consists of four NVIDIA A100 GPUs. We focus on the GPU metrics from each node, particularly the GPU temperatures.
Here we assume that we have already collected the data of size $5,824\!\times\!16,329$ (i.e., $\sim24$ hours).
Then, we incrementally add $5,825$ time points.
While the incremental additions of $5,825$ time points are completed in $29.945$ seconds, the update without using our incremental algorithm is finished in $59.263$ seconds.
We set the \texttt{max\_levels}=$9$ in each case. Both mrDMD and I-mrDMD extracted larger number of modes in the GPU metrics case, when compared to the environment logs in the previous case. This is because we increased the levels from $8$-$9$. Note that the computation time increases with the increase in the modes extracted in each level and the number of levels.
These results show that updates with our algorithms are fast enough to handle a large number of time series with extremely high velocity (i.e., $3$-second interval) in real-time.

\vspace{5pt}
\noindent\textbf{Experimental evaluation with different data sizes.}
We further evaluate our algorithms with different numbers of time points using the supercomputer temperature logs (\texttt{SC Log}) and the GPU temperature data (\texttt{GPU Metrics}).
Table \ref{table:cINCtime} shows the completion times for adding new time points. We used 6 and 7 levels for \texttt{SC Log} and \texttt{GPU Metrics}, respectively.

In Table \ref{table:cINCtime}, we first process an initial set of data consisting of varying time points ($1,000-15,000$) and a fixed number ($1,000$) of time series. 
We then add $1,000$ new time points, leading to the total processed time points of $2,000-16000$, and update the existing mrDMD modes incrementally.
The completion time for the initial data fit increases significantly as time points are added, while the incremental addition has the completion time remaining roughly the same. 
The incremental addition process, which takes only about $\sim4$ seconds for \texttt{SC Log} and $8-18$ seconds for \texttt{GPU Metrics}, showcases a remarkable reduction in compute time. Without the incremental update (i.e., equivalent to an ordinary mrDMD analysis), the overall computation time would be close to the data's initial fit, leading to significant wait times. The incremental and ordinary mrDMD approach extracted more modes for the \texttt{GPU Metrics} data when compared to the \texttt{SC Log}. Note that the levels used, the modes extracted, and the sampling rate at each level (Sec.~\ref{sec:methodology_mrdmd}) also contribute to the computation time of the mrDMD and I-mrDMD methods.

\begin{table}[tb]
\scriptsize
\centering
\caption{Completion time (in seconds) of the initial data fit (Initial Fit) and the incremental addition of $1,000$ time points (Partial Fit).}
\label{table:cINCtime}
\footnotesize
\begin{tabular}[b]{lllll}
\hline
Dataset & $N$ & $T$ &  Initial Fit & Partial Fit \\
\hline
\texttt{SC Log} & 1,000 & 2,000 & 3.6210 & 3.7667 \\
\texttt{SC Log} & 1,000 & 5,000 & 5.8420 & 4.269 \\
\texttt{SC Log} & 1,000 & 10,000 & 7.6305 & 4.1844 \\
\texttt{SC Log} & 1,000 & 16,000 & 10.3958 & 4.3255 \\
\texttt{GPU Metrics} & 1,000 & 2,000 & 7.315 & 8.654 \\
\texttt{GPU Metrics} & 1,000 & 5,000 & 20.914 & 10.583 \\
\texttt{GPU Metrics} & 1,000 & 10,000 & 28.916 & 12.953\\
\texttt{GPU Metrics} & 1,000 & 16,000 & 62.800 & 18.619\\
\hline
\end{tabular}
\end{table}

\section{Use Scenarios}
\label{sec:use_scenarios}

\begin{figure}[h]
	\centering
    \includegraphics[width=\linewidth]{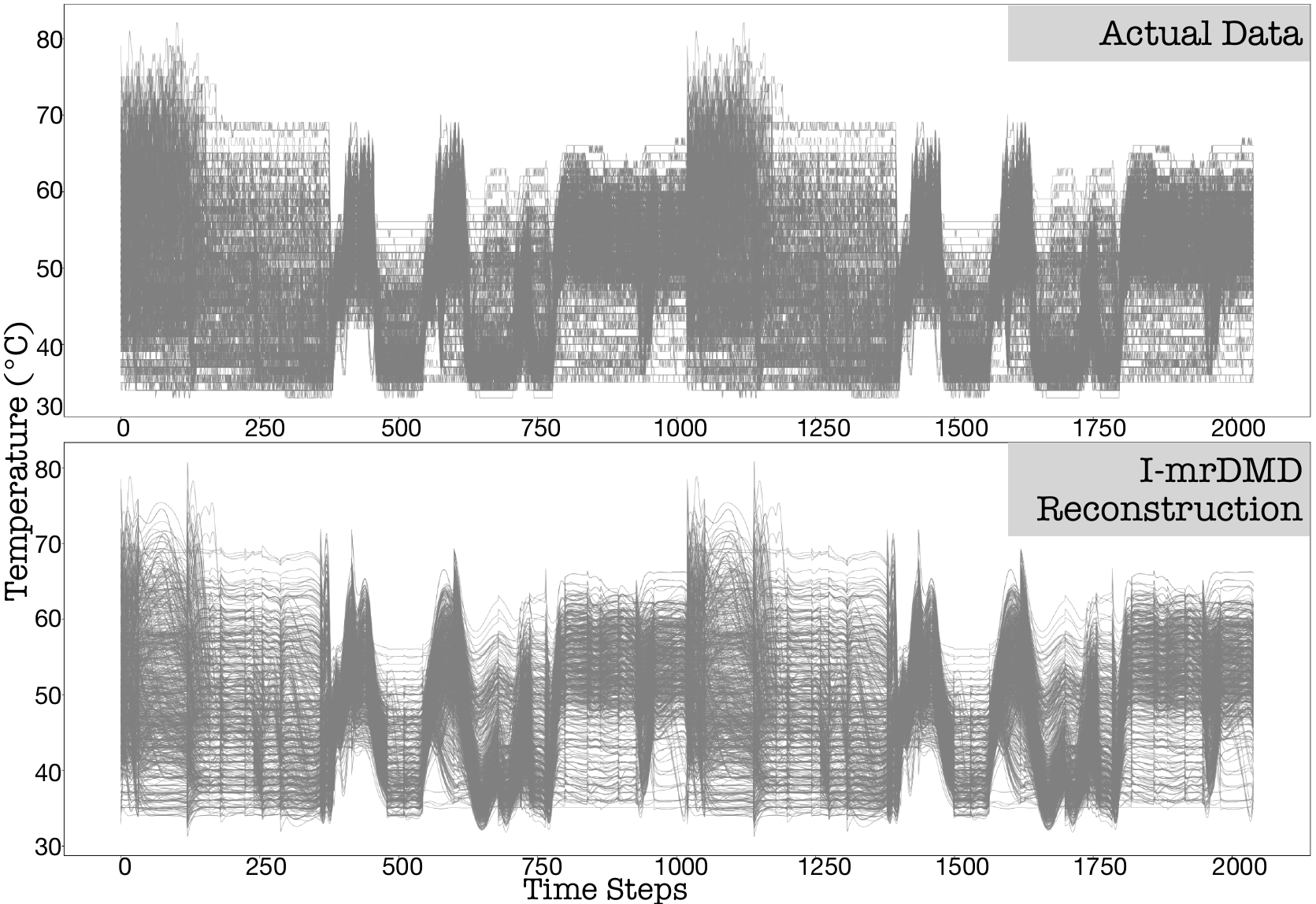}
    \caption{Analyzing the supercomputer environment logs containing baseline readings and non-baseline readings. (a) and (b) show the the actual data and the data reconstructed using the I-mrDMD modes, respectively}
	\label{cs1_1}
\end{figure}

\begin{figure*}[h]
	\centering
    \includegraphics[width=\linewidth]{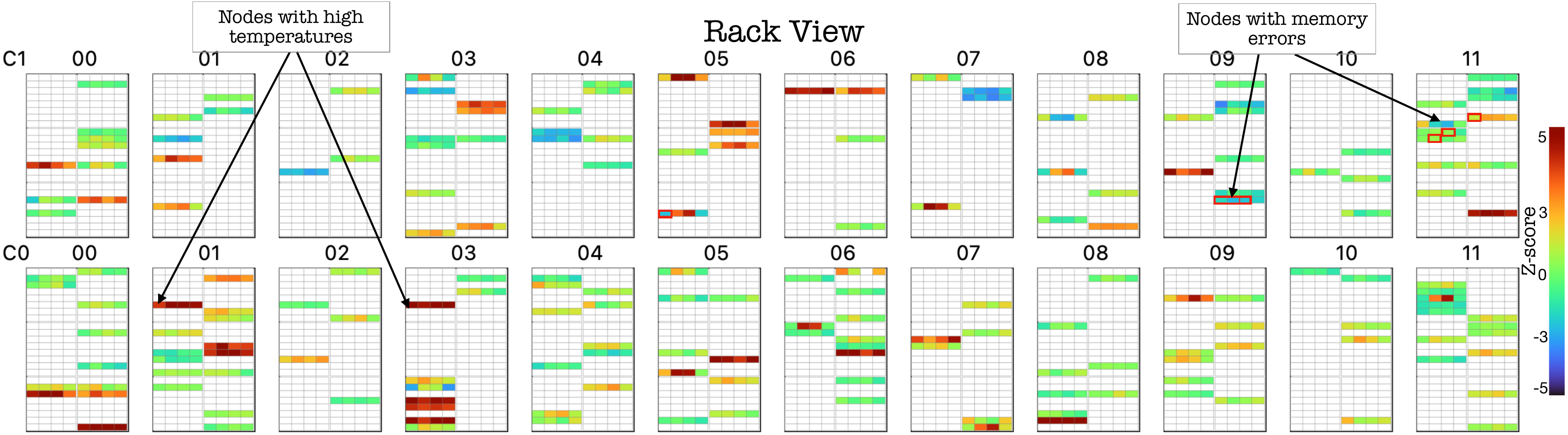}
    \caption{Figure shows the Theta supercomputer~\cite{ThetaANL} rack (or node layout) view as described in Sec.~\ref{sec:rack_vis}. The highlighted nodes are utilized by a job. The green hues represent reading closer to baseline readings, red hues represent high temperatures (much higher than baselines), and the blue hues represent low temperatures (much lower than baselines).}
	\label{cs_1_notebook}
\end{figure*}

\begin{figure}[h]
	\centering
    \includegraphics[width=\linewidth]{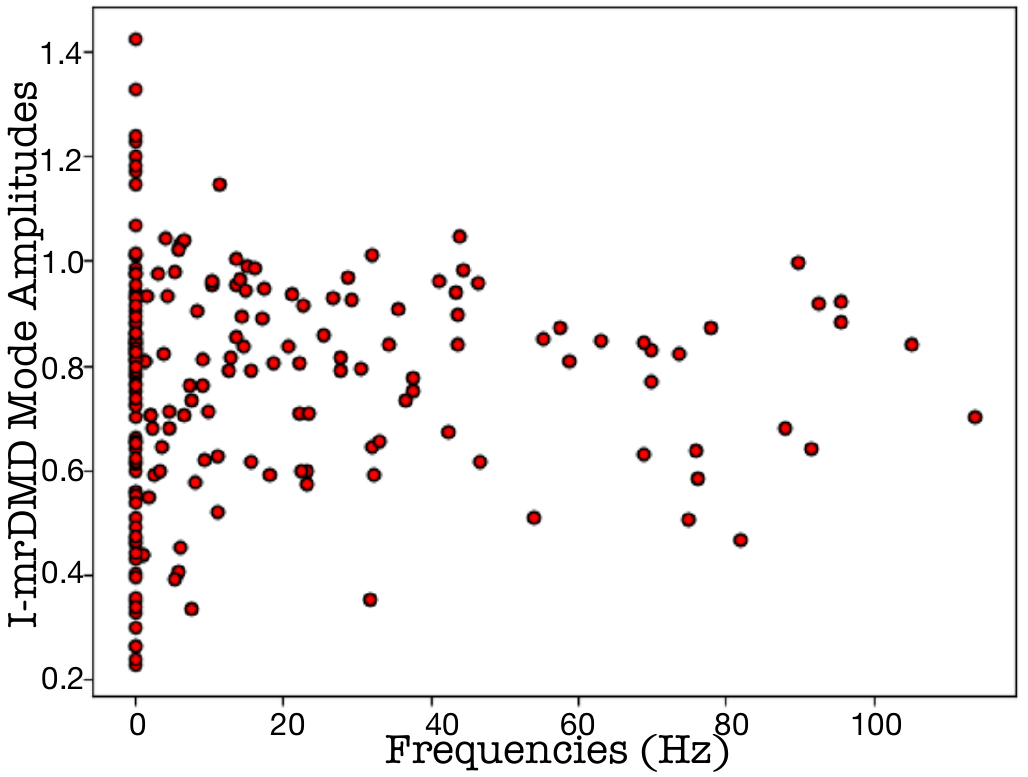}
    \caption{Figure shows mode amplitude as a function of frequency (Eq.~\ref{freq_spec}).The mrDMD spectrum shows dynamics of the data at different frequencies.}
	\label{dmd_spec}
\end{figure}

To address $Q3$, we present two in-depth case studies illustrating the utilization of supercomputers by user jobs, highlighting how our system facilitates the identification of usage patterns. These case studies showcase the results of the I-mrDMD using our D3 visualization on the Jupyter Notebook, comparing the processed data from the various previously mentioned log types.
The environment logs from the Theta supercomputer~\cite{ThetaANL} contain data from $24$ compute racks comprising $4,392$ nodes. Node sensors record data every $10$-$30$ seconds. Our primary focus is analyzing temperature readings (four readings of each type per node) from these logs. We conduct the I-mrDMD analysis and then calculate relative z-scores from baselines using the I-mrDMD results.

\subsection{Case Study 1}
\label{sec:casestudy1}
\begin{figure*}[t]
	\centering
    \includegraphics[width=\linewidth]{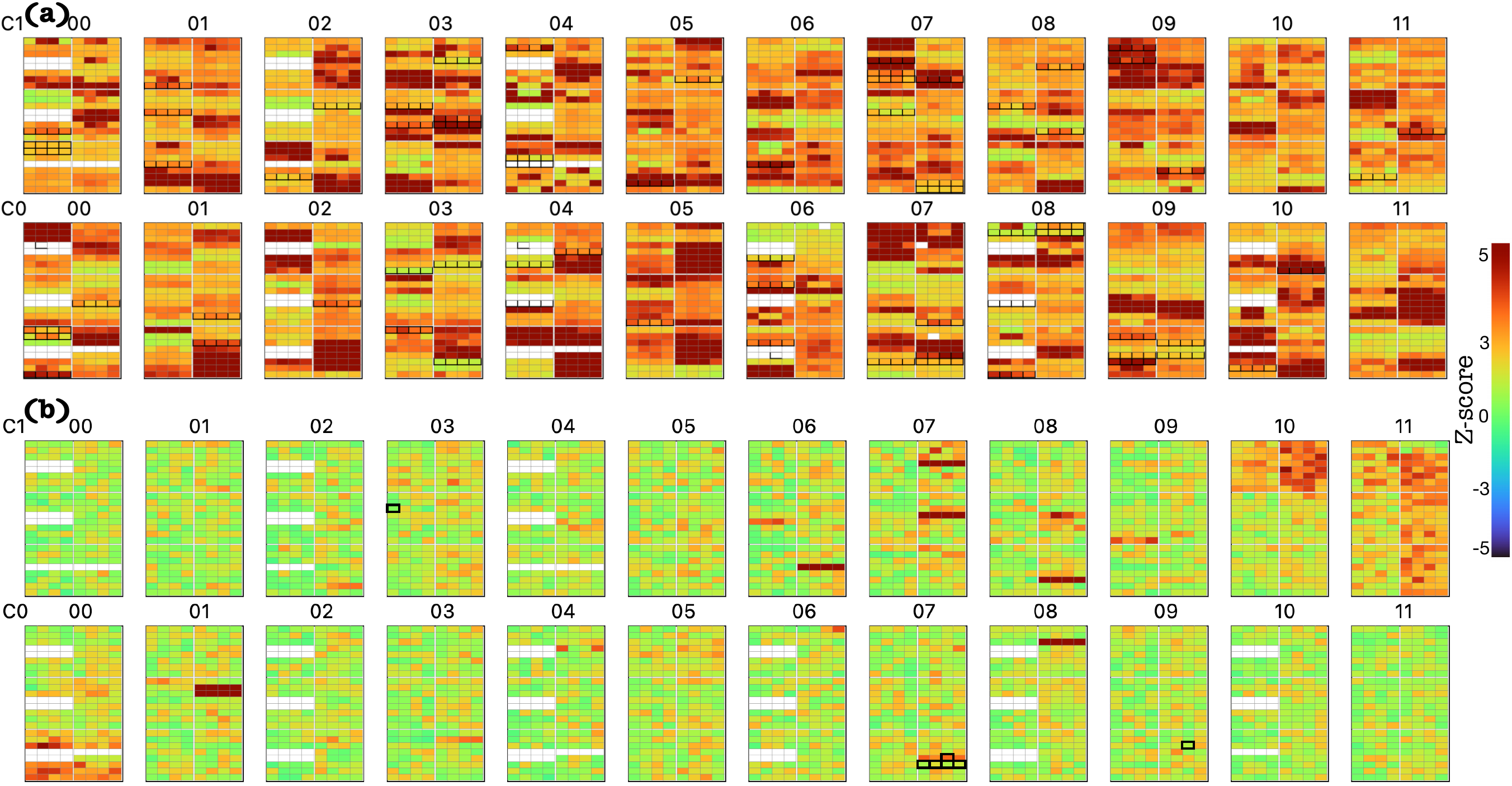}
    \caption{Figure shows the Theta supercomputer~\cite{ThetaANL} rack (or node layout) view as described in Sec.~\ref{sec:rack_vis}. We processed all nodes of the supercomputer for a period of $16$ hours with (a) showing first 8 hours and (b) showing the next 8 hours. The green hues represent reading closer to baseline readings, red hues represent high temperatures (much higher than baselines), and the blue hues represent low temperatures (much lower than baselines).}
	\label{cs_2_2}
\end{figure*}

Fig.~\ref{cs1_1} shows the time series view for 871 nodes utilized by the supercomputer. In this case study, we selected 871 nodes, which constitute nodes utilized by jobs from two projects in the facility. We first applied the mrDMD analysis on $1,000$ time steps and then incrementally updated the mrDMD results with $1,000$ additional time steps, thus simulating a streaming environment. The initial mrDMD step took $12.49$ seconds, and the incremental update took $\sim 7.6$ seconds to complete. To address $Q1$, Fig.~\ref{cs1_1}, shows the actual and the data reconstructed from the mrDMD temporal and spatial modes. We see that the reconstructed data has less high-frequency noise. We can further reduce the noise from our mrDMD analysis by selecting only high-power DMD modes from the mrDMD power spectrum and reducing the levels (refer Fig.~\ref{sys_arch}). We set the current levels to $6$ in the multiresolution step and set the frequency range to $0$-$60$Hz in the I-mrDMD spectrum. The Frobenius norm of the difference between the actual and the reconstructed data is $3958.58$. We can reduce this further by using more levels in our analysis and  including all frequencies from the mrDMD spectrum. The mrDMD spectrum for the current data is shown in Fig.~\ref{dmd_spec}. Note that increasing the levels captures the high-frequency modes in the data, and the reconstructed data would include high-frequency fluctuations.

Next, we compute the z-scores of readings of interest from the chosen baselines. The baselines are chosen so that they lie between the values $46$$^{\circ}$C$ - 57$$^{\circ}$C. Any z-score value within the $-1.5-1.5$ range is considered to be close to the baseline. Very high z-score values $>2$ constitute very high temperatures in the nodes and may cause the component to overheat and fail. On the contrary, negative z-score values represent low node temperatures, implying that the jobs are not utilizing the node and the node is possibly stalled. This scenario could result in low system utilization. We plot the results of the I-mrDMD and z-scores in the D3 visualization on Jupyter Notebook. Fig.~\ref{cs_1_notebook}, shows the node layout or rack view of the Theta supercomputer and the z-score values are shown at the respective nodes. We have used the Turbo diverging color scheme with blue hues representing negative z-scores, green representing baseline and red hues showing more positive z-scores. We see that node in close proximity show similar z-scores. The nodes highlighted in red outline are the ones showing correctable memory issues. The z-scores in these nodes are either in the negative or near-baseline range. In this case study, the elevated temperatures observed on the nodes did not indicate any hardware-related errors. However, sustained high temperatures could potentially result in overheating of the components and component degradation over time. This case indicates that utilizing results from multiple logs helps paint a bigger picture of the underlying state of these large-scale systems.

\subsection{Case Study 2}
\label{sec:casestudy2}

\begin{figure}[h]
	\centering
    \includegraphics[width=\linewidth]{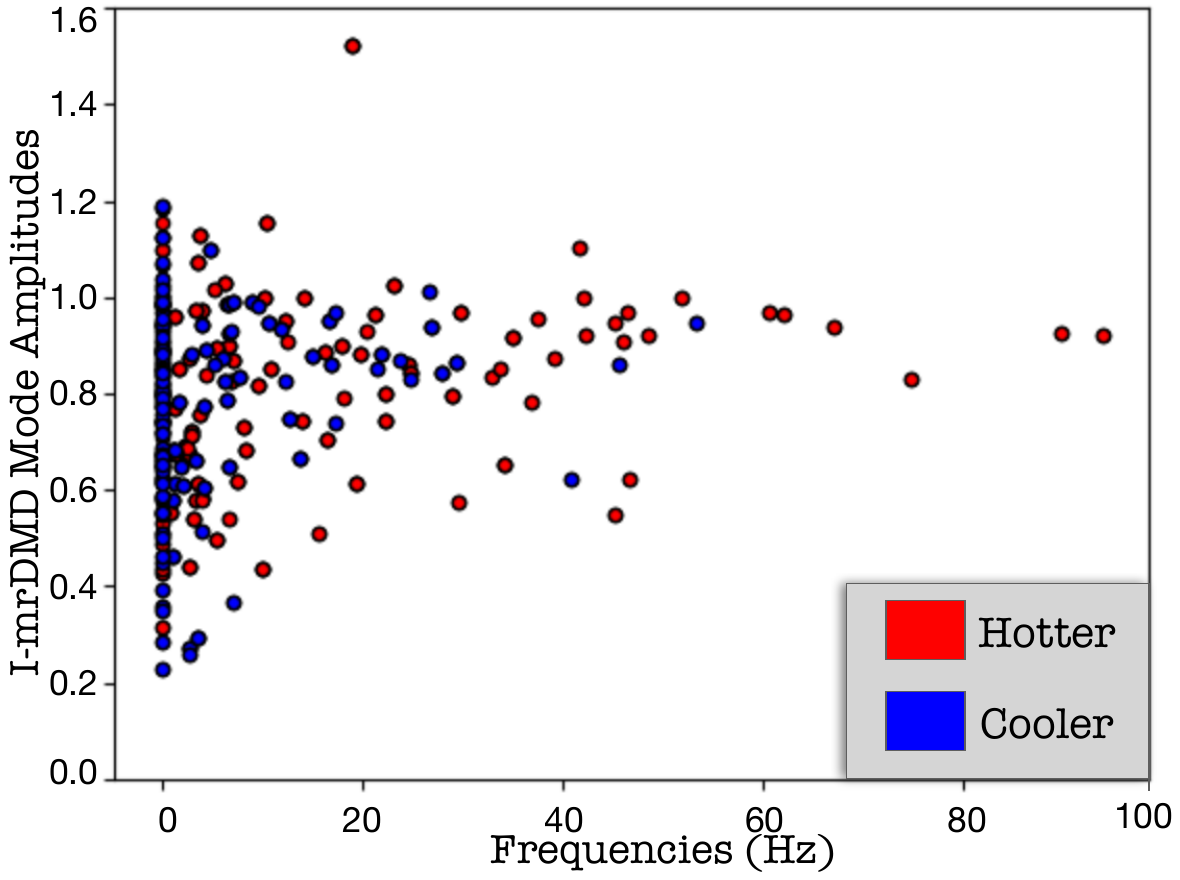}
    \caption{Figure shows mode amplitude as a function of frequency (Eq.~\ref{freq_spec}) for Fig.~\ref{cs_2_2}(b). Modes in red represent data from Fig.~\ref{cs_2_2}(a) and blue represent data from Fig.~\ref{cs_2_2}(b). Note that the baselines are different in each case and are chosen relative to the system state at time of data collection.}
	\label{cs_2_2_spec}
\end{figure}

\begin{figure*}[h]
	\centering
    \includegraphics[width=\linewidth]{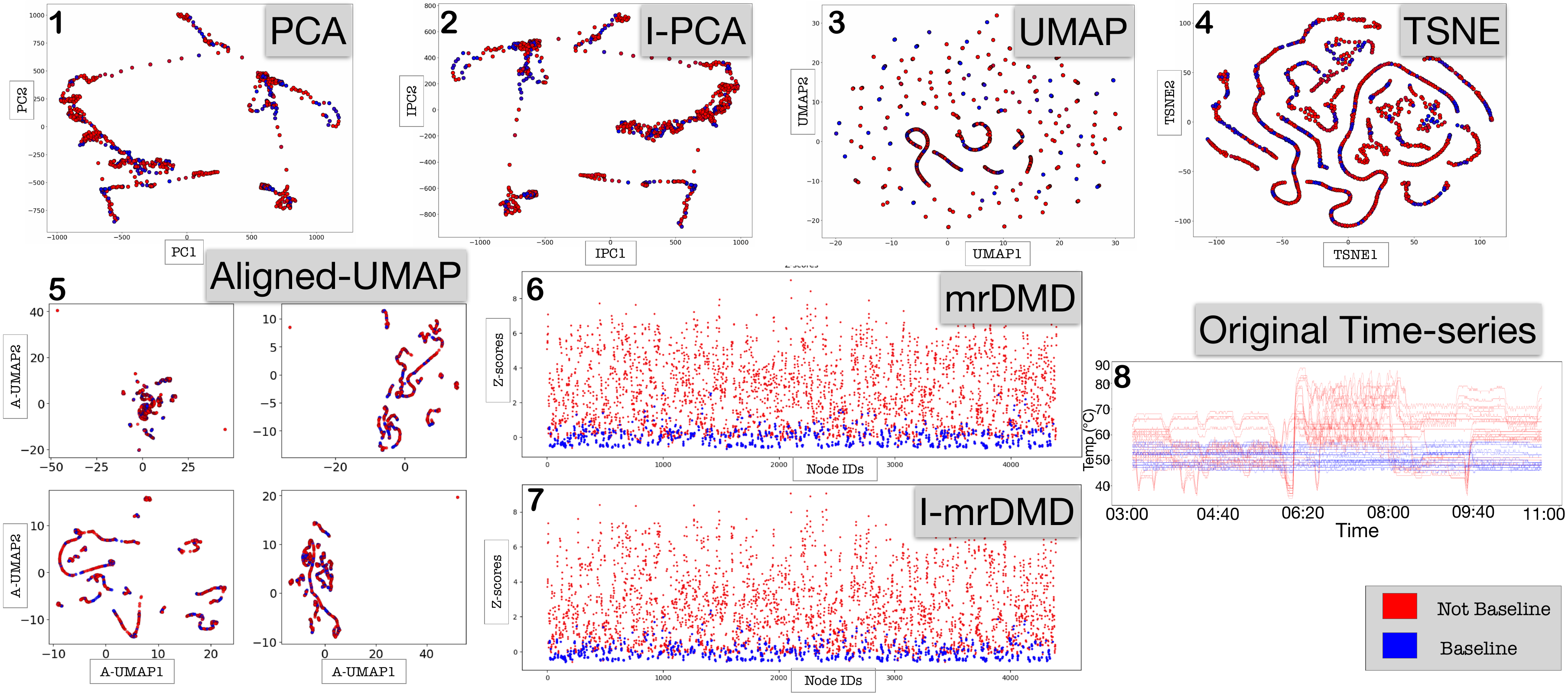}
    \caption{ The comparison of results between the (1) \texttt{PCA}, (2) \texttt{IPCA} (incremental PCA), (3) \texttt{UMAP}, (4) \texttt{TSNE}, (5) \texttt{Aligned-UMAP}, (6) \texttt{mrDMD}, and (7) \texttt{I-mrDMD} (incremental mrDMD), for the (8) original timeseries. The blue readings represent baselines and the red non-baseline readings. The figure shows $40$ ($20$ for each type: baselines and non-baselines) readings of the $4,392$ processed measurements.}
	\label{perf2}
\end{figure*}

In this case study, we selected $4,392$ nodes, which constitute nodes utilized by jobs for the duration of $8$ hours (refer Fig.~\ref{cs_2_2}(a)(b)), $16$ hours in total. We first applied the mrDMD analysis on the first $7$ hours and then incrementally updated the mrDMD results in $1,000$ time steps increments, thus simulating a streaming environment. The initial mrDMD step took 21.120 seconds, and the incremental update took $\sim$20.452 seconds. We set the number of levels to seven and split the timeline into halves at each level. The Frobenius norm of the difference between the actual and the reconstructed data is 3423.847. As mentioned in case study 1, we can further minimize reconstruction differences by using additional levels in our analysis and including all frequencies from the mrDMD spectrum.  

We compute the z-scores of the readings of interest from the chosen baselines. A z-score within the range of $-1.5$ to $1.5$ is considered near the baseline. Z-scores exceeding $2$ indicate significantly high node temperatures, which may lead to component overheating and potential failure. Conversely, negative z-scores suggest low node temperatures, indicating that the jobs are not utilizing the node and may be idle, which could lead to suboptimal system utilization. We plot the results of the I-mrDMD and z-scores in the D3 visualization on Jupyter Notebook. Fig.~\ref{cs_2_2} shows the node layout of the Theta supercomputer with the z-score values at the respective nodes. We used the Turbo diverging color scheme, where blue hues represent negative z-scores, green indicates the baseline, and red hues signify increasingly positive z-scores. Fig.~\ref{cs_2_2}(a) and (b) represent the data from the first $8$ hours and the next $8$ hours, respectively. Fig.~\ref{cs_2_2}(a) shows the significantly higher temperatures from the baselines. 
The z-scores in Fig~\ref{cs_2_2}(a) and Fig~\ref{cs_2_2}(b) are computed using different sets of baselines. For this case study, we chose these baselines from each dataset. Baselines in Fig.~\ref{cs_2_2}(a) (showing a \textit{hotter} state of the supercomputer) are selected between $45^{\circ}$C and $60^{\circ}$C, while baselines in Fig.~\ref{cs_2_2}(b) (showing a relatively \textit{cooler} state of the supercomputer) are temperatures between $30^{\circ}$C and $45^{\circ}$C. In doing so, we process the state of nodes relative to the supercomputer state at the time of processing. The users can easily customize and choose the baseline of their preference. For example, a user can choose baselines specific to the user jobs that capture the jobs complex dynamics and compare it with other jobs in the system. This would result in a better understanding of how user applications and projects use the system in the large-scale computing facility~\cite{shilpika2023multilevelmultiscalevisualanalytics}. 

Fig.~\ref{cs_2_2_spec} shows the I-mrDMD spectrum for the current data; we see that the I-mrDMD modes lie between the frequency range of $0$-$100$Hz. The blue color representing the cooler state of the system in Fig.~\ref{cs_2_2}(b) shows mode magnitudes in the lower frequency range, while the hotter system in Fig.~\ref{cs_2_2}(a) shows mode magnitudes in the higher frequency range.    
This method gives us an insight into the behavior underlying the system across different time ranges and multiple user jobs. 
In Fig.~\ref{cs_2_2}(a)-(b), the darker black node outlines denote hardware errors in the system. We see that as the day progresses, the system utilization changes (Fig.~\ref{cs_2_2}(b)), and node temperatures are closer to the baseline measurements. The node outlines in Fig.~\ref{cs_2_2}(b) highlight the nodes that persistently report hardware errors, even with multiple jobs running on the system. This suggests a potential underlying issue with these nodes that may require attention.

\section{Discussion and Future Work}
\label{sec:discussion}

\begin{figure}[h]
	\centering
    \includegraphics[width=\linewidth]{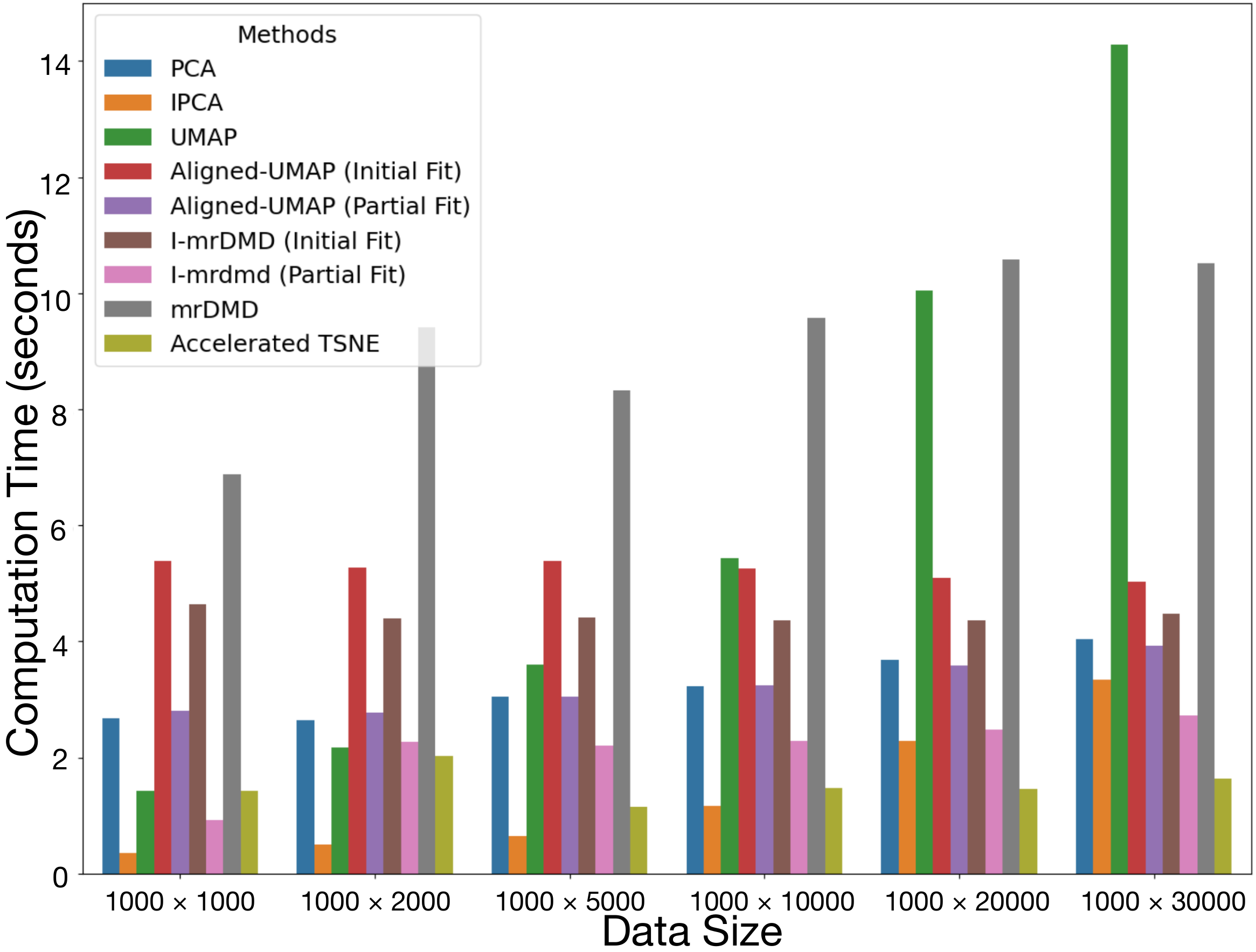}
    \caption{The comparison of completion time, showing how performance
scales with datasize.}
	\label{perf}
\end{figure}

We have demonstrated the effectiveness of our approach in analyzing large-scale time series data, highlighting its relevance for the timely processing of supercomputer environment logs to extract meaningful insights into system functionality. In this section, we further elaborate on the strengths of our approach, using the Theta supercomputer temperature readings~\cite{ThetaANL}.

In Fig.~\ref{perf}, we use the Theta environment logs,  to compare the completion time of methods and their streaming or online counterparts. 
We use the MacBook Pro (16-inch, 2023) with an Apple M2 Max chip, 32 GB memory, and OS version 14.5 for this evaluation.
We average completion times over $10$ executions.
We scale the data starting with a data size of $P\!\times\!T \!=\!1,000\!\times\!1,000$ to $1,000\!\times\!30,000$.
We use scikit-learn's implementation for \texttt{PCA} ($\mathrm{n\_components}$=$2$, $\mathrm{svd\_solver}$=$`auto$')~\cite{MACKIEWICZ1993303}, \texttt{IPCA} ($\mathrm{n\_components}$=$2$, $\mathrm{batch\_size}$=$10$)~\cite{ross2008incremental}, and \texttt{TSNE} (not shown in figure) ($\mathrm{n\_components}$=$2$, $\mathrm{learning\_rate}$=$0.01$, and $\mathrm{perplexity}$=$30$)~\cite{JMLR:v15:vandermaaten14a}, McInnes et al.'s implementation for \texttt{UMAP}~\cite{mcinnes2018umap} ($\mathrm{n\_neighbors}$=$15$, $\mathrm{n\_components}$=$2$,$\mathrm{min\_dist}$=$0.1$, and $\mathrm{metric}$=``euclidean''), and Dadu et al.'s \texttt{Aligned-Umap}~\cite{DADU2023100741} (for sequential data) ($\mathrm{n\_neighbors}$=$15$, $\mathrm{n\_components}$=$2$,$\mathrm{min\_dist}$=$0.1$, and $\mathrm{metric}$=``euclidean''). \texttt{I-mrDMD} ($\mathrm{max\_levels}$=$4$, $\mathrm{max\_cycles}$=$2$, and $\mathrm{do\_svht}$=True) and \texttt{Aligned-Umap} provide methods that incrementally update the results, called \textit{partial fit}, following the \textit{initial fit}. First, we process $1,000$ time points with the \textit{initial fit}. Next, we use \textit{partial fit} containing $1,000$ time points for the rest. The \textit{partial fit} is the result for each incremental update. Since we update the precomputed results at each $1,000$ time point updates, the \texttt{I-mrDMD} always outperforms the \texttt{mrDMD} approach, saving significant resource time. \texttt{I-mrDMD} also outperforms the \texttt{Aligned-Umap} at both \textit{initial fit} and  \textit{partial fit}. Although the \texttt{I-mrDMD} \textit{partial fit} outperforms \texttt{PCA} in compute time, it is slower than the \texttt{IPCA}. We do not show results for Multicore-TSNE~\cite{Ulyanov2016}, a batch-processing counterpart of \texttt{TSNE} because we could not install the implementation in our current system. Similar to \texttt{IPCA}, \texttt{Accelerated TSNE} outperformed \texttt{I-mrDMD}.

Fig.~\ref{perf2} shows the comparison of results generated by the methods such as (1) \texttt{PCA}, (2) \texttt{IPCA} (incremental PCA), (3) \texttt{UMAP}, (4) \texttt{TSNE}, (5) \texttt{Aligned-UMAP}, (6) \texttt{mrDMD}, and (7) \texttt{I-mrDMD} (incremental mrDMD) for the (8) original time series. The blue readings represent baselines and the red non-baseline readings. The $x$ and $y$ axes show the corresponding dimensionality reduction method components. Fig.~\ref{perf2}(8) shows $40$ ($20$ for each type: baseline and non-baseline) readings of the total $4,392$ processed measurements. We used the same settings to process the results, as mentioned previously, except we use $\mathrm{n\_neighbor}$=$400$ for \texttt{UMAP}. 
Fig.~\ref{perf2}(8) shows a separation in the behavior of baseline and non-baseline readings. Other methods, such as \texttt{PCA}, \texttt{IPA}, \texttt{UMAP}, \texttt{TSNE}, and \texttt{Aligned-UMAP}, show microclusters of non-baseline and baseline measurements grouped together. mrDMD and I-mrDMD show a separation of these measurements. Note that there is a slight variation in the results for the mrDMD and I-mrDMD as mentioned previously in Sec.~\ref{sec:use_scenarios}. We see a few non-baseline measurements in the baseline z-score range (-$1.5$-$1.5$). This behavior is because, in the current example, the dataset has very close lying measurements between the baselines and non-baselines (Fig.~\ref{perf2}(8)). If the demarcation between these measurements is clear, then the z-score distance is also larger~\cite{shilpika2023multilevelmultiscalevisualanalytics}. Here we chose a simple example for the baselines to identify how different methods capture the dynamics of the data. Baselines that capture the system's complex dynamics can be selected. This will lead to a deeper understanding of how user applications and projects utilize the supercomputer~\cite{shilpika2023multilevelmultiscalevisualanalytics}, when compared to results obtained by methods such as averaging temperatures over time.

As shown in this paper,  I-mrDMD helps save precious compute time, leading to fast analysis of large-scale computing sensor readings. Adding z-score analysis to the pipeline leads to seamless integration of the results into visualizations of the supercomputer layout, leading to an intuitive design that opens a window into the health of the supercomputing system for fast diagnosis and recovery (Sec.~\ref{sec:use_scenarios}). Since mrDMD is a data-driven approach, this methodology is generalizable and portable across multiple supercomputers. The updates to the precomputed results in the I-mrDMD levels $2$-L (refer~\ref{sys_arch}(c)), L being the max\_levels, are left as a part of future work. Users could efficiently perform these updates through asynchronous analysis for datasets that require updates. In our analysis of supercomputer logs, we found that the reconstruction of the data using I-mrDMD modes without this update was sufficient to capture the underlying dynamics of the system. The sampling rate in our I-mrDMD approach varies by level, with lower levels employing lower sampling rates and higher levels sampling data at increased frequencies. This strategy of extracting high-frequency modes at higher levels enables us to update only the slower (lower-level) modes of the precomputed results without much impact on the accuracy. 
As a part of the future work, we plan to implement this methodology on larger datasets, data collected from exascale and zettascale systems, to extract annual usage patterns. Since the current methodology adds errors in small increments, we plan to improve the implementation to eliminate these errors and include a thorough evaluation of these incremental errors on accuracy in exascale and zettascale systems. We plan to improve the divergence issue inherent in mrDMD as the temporal resolution is increased~\cite{GONZALES2022107718}. We also want to extend the I-mrDMD approach to add new entire time series or sensor measurements incrementally and perform a thorough evaluation on compression and runtime savings.

\section{Conclusion}

Our work aims to build a holistic analysis pipeline for reactive and prompt processing of massive monitoring data, particularly the environment logs. We then visually align the computed results with the hardware logs and job logs, which are collected from disparate subsystems and components of a supercomputer system. Using D3 visualization with Jupyter Notebook allows for an accessible overview of the system's underlying state and behaviors without the hassle of installing a visual analytics tool. With a provided set of supercomputer layout details, our rack layout visualization enables users to display multiple supercomputers within the notebook interface.
The primary objective of our work is to consolidate and analyze diverse types of log data. Environmental log data, often overlooked due to its massive size and the lack of efficient pattern detection mechanisms, is a key focus of our study. We aim to address this gap with our incremental mrDMD (I-mrDMD) analysis and visualization techniques. By enhancing the multiresolution dynamic mode decomposition (mrDMD) algorithm, we can swiftly identify usage and error patterns in supercomputers across different temporal and spatial scales. This data-driven approach is not only effective but also adaptable, as it can be readily applied to other large-scale system log data to uncover usage patterns in supercomputers. We plan to extend the I-mrDMD approach to add new entire time series or sensor
measurements incrementally.

\section*{Acknowledgment}
This research was funded by and used resources of the Argonne Leadership Computing Facility, a U.S. Department of Energy (DOE) Office of Science user facility at Argonne National Laboratory and is based on research supported by the U.S. DOE Office of Science-Advanced Scientific Computing Research Program, under Contract No. DE-AC02-06CH11357. We thank the ALCF HPC Systems teams and members, Operations and Network Administration, including Filippo Simini, Doug Waldron, Michael Zhang, Bill Allcock, and Ben Lenard, for their help in procuring the data and guidance in comprehending various subsystems and features.

\bibliographystyle{IEEEtran}
\bibliography{07_main}

\end{document}